\documentclass[11pt]{article}
\pdfoutput=1
\usepackage{setspace,braket}
\usepackage{amsmath, amssymb}
\usepackage{etoolbox}
\usepackage{jheppub}

    \makeatletter
    \patchcmd{\maketitle}{\@fpheader}{}{}{}
\makeatother

\newcommand{\TT}{\mathbb{T}}

\def\hsf{\mathfrak{hss}}
\def\hs{\mathfrak{hs}}
\def\hsr{\mathfrak{hsr}}

\def\be{\begin{equation}}
\def\ee{\end{equation}}

\def\R{{\mathcal R}}

\def\W{{\mathcal W}}
\def\L{{\mathcal L}}

\def\L{{\mathcal L}}

\def\be{\begin{equation}}
\def\ee{\end{equation}}

\def\e{\epsilon}

\def\l{\lambda}
\def\l{\lambda}

\def\bg{\bar{g}}

\def\beq{\begin{eqnarray}}\def\eeq{\end{eqnarray}}
\def\ba#1\ea{\begin{align}#1\end{align}}
\def\bg#1\eg{\begin{gather}#1\end{gather}}
\def\bm#1\em{\begin{multline}#1\end{multline}}
\def\bmd#1\emd{\begin{multlined}#1\end{multlined}}

\def\c{\chi}

\def\e{\epsilon}

\def\l{\lambda}
\def\L{\Lambda}

\def\p{\phi}
\def\tp{\tilde \phi}

\def\P{\Phi}

\def\pa{\partial}

\def\nn{\nonumber}

\def\({\left(}
\def\){\right)}
\def\[{\left[}
\def\]{\right]}

\newcommand{\equ}[1]{Eq.~(\ref{#1})}
\newcommand{\sct}[1]{Sec.~(\ref{#1})}

\begin{document}
\title{The Higher Spin Rectangle}
\author[a,b]{Menika Sharma}
\affiliation[a]{Department of Mathematics, City, University of London,
Northampton Square, EC1V 0HB London, UK}
\affiliation[b]{ Harish-Chandra Research Institute, Chhatnag Road, Jhusi, Allahabad 211019, India}
\emailAdd{menika.sharma@city.ac.uk}
\begin{abstract}{The chiral algebra of the symmetric product orbifold of a single-boson CFT corresponds to a ``higher spin square'' algebra in the large $N$ limit. In this note, we show that a symmetrized collection of $N$ bosons defines a similar structure that we refer to as the higher spin rectangle algebra. We explore the relation of this algebra to the higher spin square algebra. The existence of such a truncated algebra hints at bulk theories interpolating between Vasiliev higher spin theory and string theory. }
\end{abstract}
\toccontinuoustrue
\maketitle
\allowdisplaybreaks[1]

\section{Introduction}
It has long been expected that a large symmetry underlies string theory and that this symmetry is manifest in its tensionless phase. Recently, the papers \cite{Gaberdiel:2015a,Gaberdiel:2015b,Gaberdiel:2014a} endeavored to unmask this symmetry via the AdS/CFT correspondence. The CFT dual to tensionless string theory on $\textrm{AdS}_3\times \textrm{S}^3\times \textrm{T}^4$ is believed to be the symmetric product orbifold theory consisting of free fermions and bosons on $({\TT}^4)^N/S_N$. In Ref.~\cite{Gaberdiel:2015a} it was shown that, in the large $N$ limit, the chiral algebra of the single-trace operators for this symmetric product orbifold can be organized in terms of $\W$-algebra representations. Using the bulk/boundary dictionary \cite{Gaberdiel:2010,Gaberdiel:2011,Gaberdiel:2012}, in this case, leads to an infinite-dimensional lie algebra as the symmetry algebra of string theory on the AdS background at the tensionless point. The associated gauge fields are towers of massless higher-spin fields, with the gauge fields of Vasiliev theory \cite{vas1,vas2} appearing as a sub-sector. Because the generators of this lie algebra can be organized in a twofold way in terms of the representations of the higher spin algebras underlying Vasiliev theory, it was christened the higher spin square ($\hsf$). Recent work \cite{Gaberdiel:2017} has attempted to directly identify the generators of this higher spin square algebra with massless states in tensionless string theory on the $\textrm{AdS}_3$ background. Regardless of its status as the symmetry algebra of string theory, however, the $\hsf$ algebra as well as the full chiral algebra of the symmetric product orbifold theory, and its relation to conventional $\W$-algebras, are interesting objects to study in their own right.

$\W$-algebras have a long history going back to the work of \cite{Zamolodchikov:1985wn} where they appeared as the symmetry algebra of parafermionic conformal field theories. As such they are generalizations of the Virasoro algebra, with the $\W_N$ algebra defined as an algebra with generators having spin $2$ to $N$. In the context of string theory, historically, parafermionic models have been used to formulate world-sheet theories \cite{Argyres:1991,Pope:1991} with $\W$-algebras replacing the Virasoro algebra. 

In recent times, the study of $\W$-algebras has been revitalized because of their identification as the asymptotic symmetry algebras of three-dimensional Vasiliev theories \cite{Campoleoni:2010}. In particular, the $\W_\infty[\lambda]$ algebra is the asymptotic symmetry algebra of a Vasiliev theory with bulk gauge group $\hs[\lambda]$, where the parameter $\lambda$ is related to the bulk scalar mass. In fact, the $\W$-algebras do not appear in isolation but are deeply connected to lie algebras \cite{Bouwknegt:1992}. The $\W(\mathfrak L)$ algebra, where $\mathfrak L$ denotes a lie algebra, is the Drinfel'd-Sokolov reduction of the affine algebra based on $\mathfrak L$ \cite{Bershadsky:1989}. Thus, $\W_N$ can be constructed from $\mathfrak{sl}(N)$ while $\W_\infty[\lambda]$ can be constructed from $\hs[\lambda]$. A reverse construction also exists. The algebra $\W_\infty[\lambda]$ is actually a two-parameter algebra with one parameter being $\lambda$ and the second being the central charge $c$. One can recover the lie algebra $\mathfrak L$ from $\W(\mathfrak L)$ by the Bowcock-Watts procedure~\cite{Bowcock:1991, Gaberdiel:2011wb}, which consists in taking the central charge to infinity while also restricting the operators such that the absolute value of their mode number is less then the spin, {\it i.e.} to the so-called wedge modes. The algebra $\mathfrak L$ is, therefore, often referred to as the wedge sub-algebra of $\W(\mathfrak L)$. In this notation, the $\hsf$ algebra is the wedge algebra of the chiral algebra of the large $N$ symmetric product orbifold CFT.

There are also various relations between different $\W$-algebras.  For example, for $\W_\infty[\lambda]$ changing the value of the parameter $\lambda$ changes the algebra: $\W_\infty[0]$ is a linear algebra while generically $\W_\infty[\l]$ is a non-linear algebra. However, there exists a triality symmetry \cite{Gaberdiel:2012ku} --- the algebras are equivalent for generically three different values of $\lambda$. The algebra $\W_\infty[\lambda]$  truncates to $\W_N$ for $\lambda=N$ and a specific value of the central charge. The algebra $\W_\infty[\lambda]$ therefore acts as a ``master algebra'' in the sense that the finite $\W$-algebras can be extracted from it. Similarly, for a given $\lambda$, changing the value of $c$ can lead to a representation of the $\W_\infty[\lambda]$ algebra developing null states such that it truncates to a representation of a smaller $\W_N$ algebra. 

The algebra $\W_\infty[\lambda]$ also has an infinite-dimensional sub-algebra $\W^e_\infty[\lambda]$, which consists of fields of even spin only. This algebra, its supersymmetric extensions and holographic duals have been explored in \cite{Candu}. In the other direction, $\W_\infty[\lambda]$, as also $\W_N$ algebras, can be extended by adding a spin one field: the resulting algebra is known as $\W_{1+\infty}[\lambda]$.

Since the higher spin square algebra is exponentially larger than the $\W$-algebras, one might expect that a truncation of the higher spin square algebra leading to a smaller algebra exists and in fact there may exist several different kind of truncations. It is the goal of this paper to construct a realization of one such truncated algebra. The operators of the full asymptotic symmetry algebra associated with the bosonic version of the $\hsf$ can be organized in an infinite number of representations of the $\W^e_\infty[1]$ algebra interpreted one way or in an infinite number of representations of the $\W_{1+\infty}[0]$ algebra interpreted the second way. The operators, thus, cover a ``square'' with $\W_{1+\infty}[0]$ operators in the horizontal direction and $\W^e_\infty[1]$  operators in the vertical direction.
We show that truncating the operators of the $\hsf$ algebra such that the operators now organize in a finite number of $\W^e_\infty[1]$ representations still leads to a closed algebra. This set of operators can also be organized in an infinite number of representations of the $\W_{1+N}$ algebra, thus retaining the structure of the original algebra. Because we have truncated the $\hsf$ algebra in one direction only, we use the notation the higher spin rectangle ($\hsr$) algebra to refer to the wedge algebra of this truncated algebra.

While the AdS/CFT correspondence has proved remarkably potent at large $N$, it has had limited success at finite $N$. It is, thus, important to look for scenarios where the quantum version of this duality can be verified. This is another motivation, in the larger context, to catalogue and understand finite versions of the $\hsf$ algebra. In this paper, however, we will construct the $\hsr$ algebra at infinite central charge with only a few observations about the truncation of the $\hsf$ at finite central charge. 

This paper is organized as follows. We first review the construction of the higher spin square algebra in \sct{Review}. In \sct{main} we construct a realization of the $\hsr$ algebra in terms of free bosons with background charges. In \sct{Identities}, we show that this construction is reflected in $q$-series identities. In \sct{finiteN} we discuss the possibility of a finite central charge version of the $\hsf$ algebra. In \sct{dis} we discuss various directions for further exploration.

\section{Review of the higher spin square algebra \label{Review}}

In this paper, we will confine our analysis to bosonic theories as these capture all the essential details of the $\hsf$ algebra. As in Ref.~\cite{Gaberdiel:2015a}, we start with analyzing the symmetry algebra of the $N$'th symmetric orbifold of a single real boson, where we will take $N$ to $\infty$.  In the text, we will use the notation $\R \W[\infty]$ to refer to this bosonic chiral algebra. 

In the large $N$ limit, the ``single-particle'' generators for the symmetric product orbifold are symmetrized products of the form
\be\label{spgen}
\sum_{i=1}^{N} \, (\partial^{m_1} \phi_i) \cdots (\partial^{m_p} \phi_i ) \ , \qquad 
m_1,\ldots, m_p\geq 1 \ .
\ee
Because of the symmetrization over $N$, this set of generators is in one-to-one correspondence with the chiral sector of a single boson. Removing the terms that are total derivatives, and in the $N\rightarrow \infty$ limit, they also constitute a set of linearly independent operators.

The aim of Ref.~\cite{Gaberdiel:2015a} was to the arrange these single-particle generators of the bosonic theory in terms of higher spin subalgebras. To begin this task one notes that the subset of generators of \equ{spgen} of the form
\be\label{bgen}
\sum_{i=1}^{N}\,  (\partial^{m_1} \phi_i)\,  (\partial^{m_2} \phi_i) \ , \qquad m_1,m_2\geq 1 \ , 
\ee
define quasiprimary generators of spin $s=m_1+m_2$, in specific linear combinations and when $s$ is even. In fact, only one independent current  can be constructed at each even spin, meaning that it is not a linear combination of derivatives of lower-spin currents, and there are no independent odd-spin currents. This set of independent currents generate the even spin $\W$-algebra ${\cal W}^{e}_\infty[1]$. 

\begin{figure}[t!] 
\begin{center}
\includegraphics[scale=1]{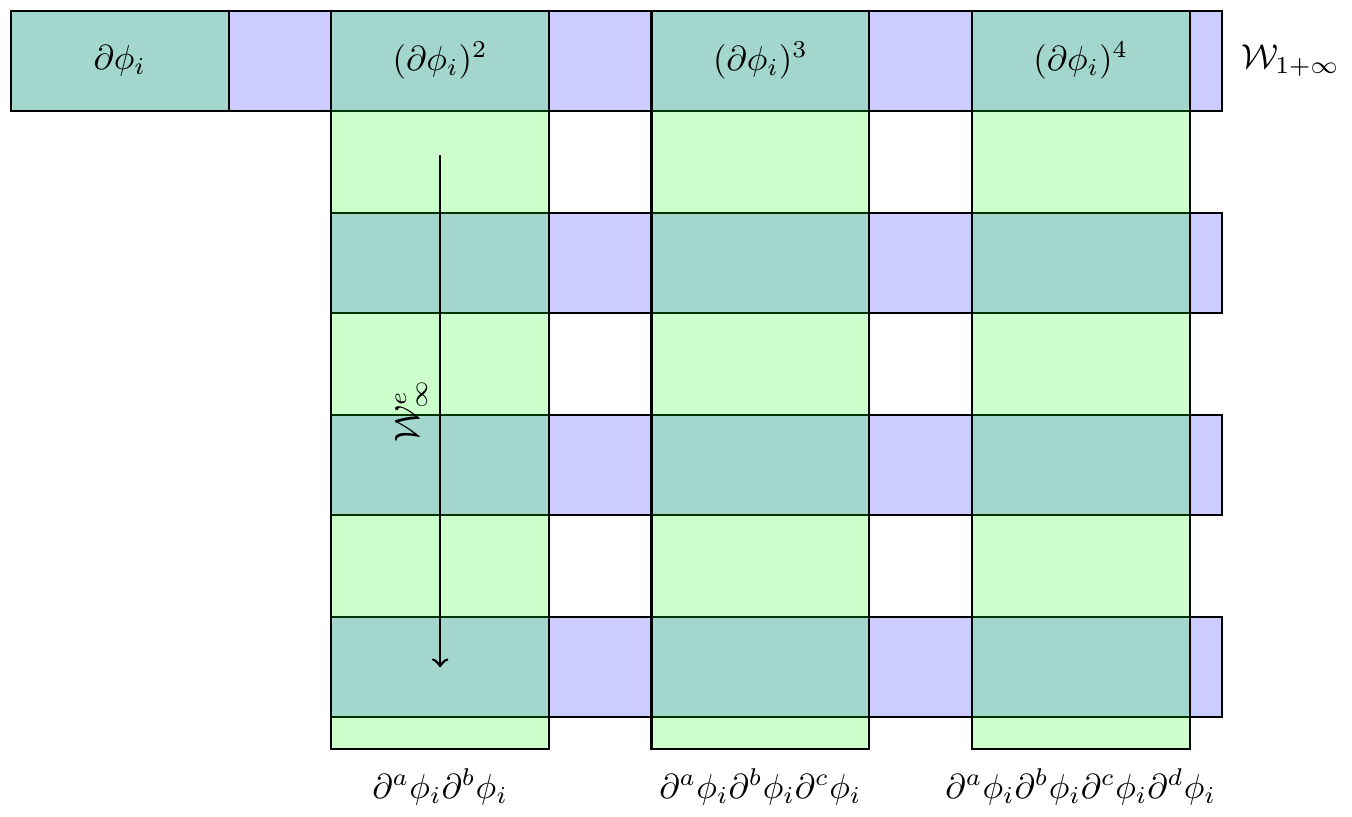}
\caption{\label{squafig} The operators in the top-most row realize the $\W_{1+\infty}[0]$ algebra. The full algebra is generated by acting by derivatives on the operators in the top row. The second column corresponds to the $\W^e_\infty[1]$ algebra while subsequent columns correspond to its representations. Operators in each row also fall into representations of $\W_{1+\infty}[0]$.}
\end{center}
\end{figure}

The generators in \equ{bgen} are of order two, i.e., they are bilinear in the $\phi$s. The currents in \equ{spgen} are of arbitrary order $p\geq 1$.  However, it turns out that the currents of a fixed order $p$, suitably corrected by lower-order terms, form a representation of the wedge algebra of ${\cal W}^{e}_\infty[1]$. This is captured in Fig.~(\ref{squafig}) where currents of a given order correspond to columns. The entire operators of the $\R \W[\infty]$  algebra are, therefore, organized into representations of ${\cal W}^{e}_\infty[1]$. 

The $\W$-algebras that we deal with in this paper are the symmetry algebras of cosets of the form $\frac{g_{k} \otimes g_{1}}{g_{k+1}}$. We use the coset notation $(\Lambda_+; \Lambda_-)$ to denote a representation of a $\W$-algebra where $\Lambda_+$ is a representation of $g_{k}$ and $\Lambda_-$ is a representation of $g_{k+1}$. Then the statement that operators of the $\R \W[\infty]$ can be organized in representations of ${\cal W}^{e}_\infty[1]$ is captured by the following identity:
\be
\label{VIfull}
	\prod_{n=1}^\infty \frac{1}{(1- q^n)} = 1+\sum_{p=1}^{\infty} b^{({\rm wedge}) [\lambda=1]}_{([0^{p-1},1,0,\ldots,0];0)} \,,
\ee
where $b^{({\rm wedge}) [\lambda=1]}_{([0^{p-1},1,0,\ldots,0];0)} $ denotes the wedge character of the $([0^{p-1},1,0,\ldots,0];0)$ representation of ${\cal W}^{e}_\infty[1]$. Specifically it is
\be
b^{({\rm wedge}) [\lambda=1]}_{([0^{p-1},1,0,\ldots,0];0)}  = \frac{q^p }{\prod_{r=1}^{p} (1-q^r)}\,.
\ee
The LHS of \equ{VIfull} is the normalized partition function for a single boson. Combinatorially, the LHS is just the generating function for the number of ways one can partition an integer varying from $1$ to $\infty$ into an arbitrary number of parts. Each term in the sum on the RHS is number of ways one can partition an integer into exactly $p$ parts. In terms of the operators in \equ{spgen}, this is the spin $s$ of the operator, varying from $p$ to $\infty$, being partitioned into $m_1,m_2,\ldots,m_p$ at fixed $p$.

One could have started sorting the operators of $\R \W[\infty]$ from a different point leading to a different organization. Operators of the form
\be\label{wgen}
\sum_{i=1}^{N}\,   (\partial \phi_i)^m \ , \qquad m \geq 1 \, ,
\ee
corrected by lower-order terms (involving smaller powers of $\partial \phi_i$ and additional derivatives) define the linear $W_{1+\infty}[0]$ algebra. This is actually a result of bosonization of the free-fermion representation of $\W_{1+\infty}$ \cite{vas2,Pope:1991,Pope:1990be, Gaberdiel:2013}. Similar to the previous case, the rest of the operators in \equ{spgen} can be organized in terms of representations of ${\cal W}_{1+\infty}[0]$. This is formalized by the identity
\be
\label{HIfull}
	\prod_{n=1}^\infty \frac{1}{(1- q^n)} =1+ \sum_{d=1}^{\infty} b^{({\rm wedge}) [\lambda=0]}_{(0;[d,0,0,\ldots,d])}\,,
\ee
where
\be
b^{({\rm wedge}) [\lambda=0]}_{(0;[d,0,0,\ldots,d])}  = \frac{q^{d^2} }{\prod_{r=1}^{d} (1-q^r)^2}\,.
\ee
Combinatorially, each term in the sum of the RHS can be interpreted via the Durfee square construction \cite{Andrews}. The Durfee square of a partition is the largest square array that can fit in the upper left corner of its Young diagram. Each term in the RHS, therefore, is the number of partitions, of an integer, having fixed Durfee number $d$. In terms of the fields, the Durfee number corresponds to the number of fermion bilinears needed to construct a particular operator.


\section{A collection of $N$ bosons and the $\hsr$ algebra \label{main}}
The $\W_{1+N}$ algebra is a non-linear algebra which has generators ranging in spin from 1 to $N$. It is related to $\W_{N}$ as
\be 
\W_{1+N} = H \oplus \W_{N} \,,
\ee
where $H$ is a spin one scalar field. It is well-known that the $\W_{1+N}$ algebra has a realization, at arbitrary central charge, in terms of $N$ free bosons and a background charge parameter \cite{Bouwknegt:1992,Prochazka:2014}. In this section we show that a symmetrized system of $N$ free bosons can be used to realize a larger algebra that we refer to as $\R\W[N]$, with its corresponding wedge algebra denoted by $\hsr [N]$. As in the case of the higher spin square algebra, this comes about because a collection of $N$ bosons can be used to represent both the $\W_{1+N}$ and the $\W^{e}_{\infty}$ algebras. In general, to construct a $\W$-algebra with $D$ independent currents, one needs a minimum of $D$ free bosons so as to have a Hilbert space large enough to accommodate the $\W$-algebra representations. At specific values of the central charge, however, $\W$-algebra representations have a large number of null vectors so that a realization of the algebra with a smaller number of bosons can be constructed.

We first review the $N$-boson realization of $\W_{1+N}$, which is commonly known as the Miura transformation, in \sct{ff1}. We also provide an alternate formulation of the Miura transformation in which the current operators are expressed as power sums. Then in \sct{ff2} we review the $N$-boson realization of the $\W^{e}_{\infty}$ algebra. In \sct{OO}, we propose that a specific set of operators that can be constructed out of $N$ free bosons, and are invariant under $S_N$ symmetry, are the generators of the $\R\W[N]$ algebra. We will show that this set of generators has closed commutation relations. This follows from the same logic as was utilized in the $N\rightarrow\infty$ case in Ref.~\cite{Gaberdiel:2015a}, which we will elaborate on in \sct{OO}. This algebra can be deformed to an arbitrary (integral) value of the central charge by changing the background charge and the total number of bosons.

\subsection{Free field representations}
\subsubsection{Free field representation of $\W_{1+N}$ \label{ff1}}
In this section we review the representation of $\mathcal{W}_{1+N}$ in terms of $N$ free bosons. This algebra has $N$ currents denoted by $\W^{(k)}(z)$, where $k$ varies from $1$ to $N$ and is the conformal dimension. These currents are defined by the generating operator
\begin{equation}
\label{miurar}
R(z) = \; : \prod_{j=1}^{N} \Big( \alpha_0 \partial + i \,\partial \phi_j(z) \Big) : = \sum_{k=0}^N \W^{(k)}(z) (\alpha_0 \partial)^{N-k}\,.
\end{equation}
The parameter $\alpha_0$  is a background charge related to the central charge of the algebra as
\be
\label{bcc}
	c = N \left\{1- \alpha_0^2(N^2-1) \right\}\,.
\ee
Let us define $J_i = i \, \partial \phi_i$. Then, for example, uptil $k=3$ the $\W^{(k)}(z)$ are given explicitly by
\begin{eqnarray}
\label{bccurrents}
\W^{(0)} & = & \mathbf{1} \,, \nn \\
\W^{(1)} & = & \sum_{j=1}^N J_i \,,\nn \\
\W^{(2)} & = &  \sum_{j<k} :J_j J_k : ~+~ \alpha_0 \sum_{j=1}^N (j-1) \,J_j^\prime \,, \nn \\
\W^{(3)} & = & \sum_{j<k<l} :J_j J_k J_l: ~+~ \alpha_0 \sum_{j<k} (j-1) :J_j^\prime J_k: \nn \\
& & ~+~ \alpha_0 \sum_{j<k} (k-2) :J_j J_k^\prime: ~+~ \frac{\alpha_0^2}{2} \sum_{j=1}^N  (j-1)(j-2)J_j ^{\prime\prime} \,. 
\end{eqnarray}
For the background charge $\alpha_0=0$ the corresponding central charge is $c=N$. Note that this central charge is equal to $1+c_\textrm{coset}$ where $c_\textrm{coset}$ is the central charge of the coset theory $\frac{SU(N)_k \otimes SU(N)_1}{SU(N)_{k+1}}$ in the limit $k \rightarrow \infty$. At this value of $\alpha_0$ the currents are of the particularly simple form
\begin{align}
\label{simpleCurrents}
\W^{(1)}(z) & = \sum_{i=1}^{N} i \, \partial   \phi_i \,, \nn \\
\W^{(2)}(z) & = - \sum_{i<j}: \partial  \phi_i  \partial \phi_j: \,, \nn \\
\W^{(3)}(z) & =-  \sum_{i<j<k} i : \partial  \phi_i  \partial  \phi_j \partial  \phi_k:\, ,\nn\\
\vdots ~~~~~&
\end{align}
Viewed as polynomials in $J_i$, the above operators correspond to the elementary symmetric polynomials in $N$ variables. The currents listed above can be thought to form an algebraically independent set of operators in the sense that the eigenvalues of the $\W^{(i)}(z)$ operators acting on a highest weight state are the elementary symmetric polynomials (which are algebraically independent) \cite{Bouwknegt:1992}. 

Using the Newton-Girard formulae, the elementary symmetric polynomials can be expressed as power sums. Analogously, for the bosonic fields we can make the following change of basis
\be
	\tp_i = \sum_{j=1}^{N} c_{i j} \phi_j\,,
\ee
where $c_{i j} = c_{i j^\prime}$ with  $j^\prime = j+ (i-1) \textrm{mod~} N$. 
Under this change of basis, the generators of the $\mathcal{W}_{1+N}$ algebra take the form: 
\begin{align}
\label{NGP}
\W^{(1)}(z) & = \sum_{i=1}^{N} i \, \partial   \tp_i \,, \nn \\
\W^{(2)}(z) & = \sum_{i=1}^{N}: \partial \tp_i  \partial  \tp_i: ~+~ a_{20} :\bigg( \sum_{i=1}^{N} \partial \tp_i \bigg)^2: \,, \nn \\
\W^{(3)}(z) & = \ \sum_{i=1}^{N} i : \partial \tp_i  \partial \tp_i  \partial \tp_i: ~+~a_{30}:\bigg( \sum_{i=1}^{N}  \partial \tp_i \bigg) \bigg(\sum_{j=1}^{N}  \partial {\tp_i}^2 \bigg): ~+~a_{31}: \bigg(\sum_{j=1}^{N}  \partial \tp_i\bigg)^3:\, \nn\\
\vdots ~~~~~&
\end{align}
where the $a_{kl}$ are constants dependent on the $c_{ij}$. This form of the currents is akin to the form in which the $\W_{1+\infty}[0]$ currents are expressed in the bosonic basis.
Indeed, the $\W_{1+N}$ algebra, like the $\W_{1+\infty}[0]$ algebra, also has a formulation in terms of complex fermions. The bosonic current $J_i(z)$ can be written in terms of a multi-component free fermion as
\be
	J_i(z) = :\overline{\psi}_i (z)\psi_i(z):\,,
\ee
where $i=1$ to $N$.
This corresponds to the bosonization
\be
	\psi_i(z) = :\exp{\{-i \phi_i(z)\}}:\,.
\ee
In the fermonic basis, the $\W$-algebra generators can be expressed in terms of bilinears of free fermions (See,{ \it e.g.}, \cite{Bershtein:2017}). There is, therefore, a smooth way of taking the limit of the Miura construction as $N\rightarrow \infty$ to give $\W_{1+\infty}[0]$ at $c=\infty$. 
\subsubsection{Free field representation of $\W^{e}_{\infty}$ \label{ff2}}
There is a well-known formulation of the $\W^{e}_{\infty}$ algebra for $\lambda=1$ in terms of $N$ free bosons \cite{Bakas:1990sh, Bakas:1990ry}. Explicitly the currents are given by
\be
	\W^{(2s)} (z) = \sum_{i=1}^{N} \sum_{k=1}^{2s-1} (-1)^k A^{2s}_k :\partial^k \tp_i \partial^{2s-k} \tp_i: \,,
\ee
where $A^{2s}_k$ is a positive number dependent on the spin. 
In this formulation the stress-energy tensor is
\be
\label{st1}
\W^{(2)} (z) = \sum_{i=1}^{N}  :\partial \tp_i \partial \tp_i: \,.
\ee
This is related to the spin~$2$ current in \equ{simpleCurrents} by a linear change of basis, as we saw in the previous section. There, thus, exists an alternate formulation of bilinear realizations of the currents in terms of $N$ bosonic fields.  We can choose the $\W^{(2)}$ current to be of the form
\be
\label{st2}
\W^{(2)}(z)  = \sum_{i<j}: \partial \phi_i  \partial \phi_j: \,.
\ee
To find the number and form of currents at each spin we proceed as usual \cite{Pope:1991} and write down all possible symmetric bilinear terms. At spin $3$ there exist the fields 
\be
	\bigg \{ \sum_{i \neq j} :\pa^2 \p_i \pa \p_j :\bigg \} {\rm~~and~~} \bigg \{\sum_{i} : \pa^2 \p_i \pa \p_i: \bigg \}
\ee
which are not independent of each other, as they are related by a change of basis. In addition they are derivatives of the spin $2$ currents given in \equ{st2} and \equ{st1} respectively. There is, therefore, no new current at spin $3$. At spin $4$, the symmetric bilinear terms are
\be
	\bigg \{\sum_{i \neq j} : \pa^3 \p_i \pa \p_j: , \sum_{ i \neq j}  :\pa^2 \p_i \pa^2 \p_j : \bigg \} {\rm~~and~~}\bigg  \{\sum_{i}  :\pa^3 \p_i \pa \p_i:, \sum_{i} :\pa^2 \p_i \pa^2 \p_i: \bigg \}\,.
\ee
The two set of currents are again not independent of each other, so that we choose only the first set, for example. In this set, the linear combination $\sum_{i \neq j} : \pa^3 \p_i \pa \p_j : +  \sum_{i \neq j}  :\pa^2 \p_i \pa^2 \p_j: $ is a derivative of $\sum_{i \neq j} :\pa^2 \p_i \pa \p_j: $. The second linear combination $\sum_{i \neq j}  :\pa^3 \p_i \pa \p_j:  -  \sum_{i \neq j}  :\pa^2 \p_i \pa^2 \p_j:$ is an independent current. Proceeding in this manner we find a single current at each even spin. 
Note that the symmetrization condition has reduced the degrees of freedom in the $N$-boson system so that the currents are in one-to-one correspondence with those that can be constructed out of a single boson. If we do not require symmetry under $S_N$, the bilinears that can be constructed from two bosons, for instance, give rise to the $\W_{\infty}$ algebra. 

The above counting is somewhat naive as it does not take into account relations that may reduce the set of independent generators at finite $c=N$.  \footnote{I thank Matthias Gaberdiel for this observation.} However, the above analysis does capture the maximal set of independent bilinear generators that can be constructed out of $N$ bosons subject to the symmetrization condition. In actual fact, a representation of the $\W^{e}_{\infty}$ algebra at finite central charge will truncate to a representation of a smaller algebra because of the presence of null fields \cite{Blumenhagen:1994}. 
\subsection{Extending the algebra \label{OO}}
\begin{figure}[t!] 
\begin{center}
\includegraphics[scale=1]{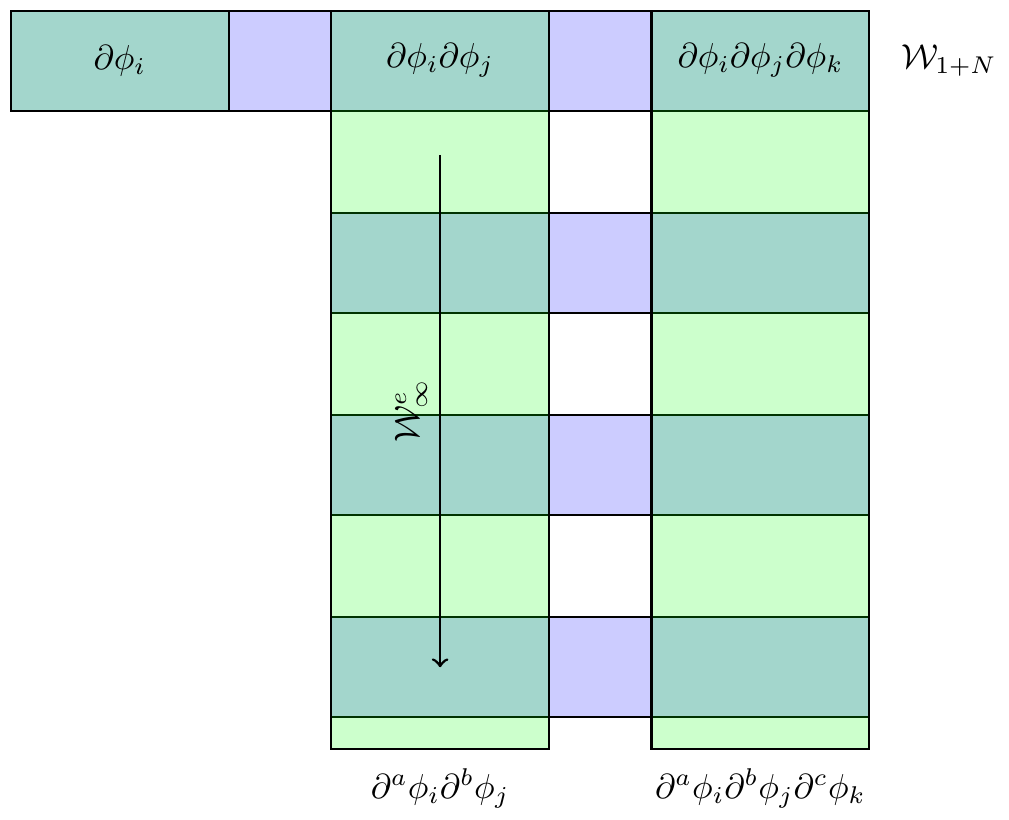}
\caption{\label{rectfig} The operators in the top-most row realize the $\W_{1+N}$ algebra. The full algebra is generated by acting by derivatives on the constituent terms of each $\W_{1+N}$ operator. The second column corresponds to the $\W^e_\infty[1]$ algebra. For ease of notation, we have dropped the summation symbol for the terms in the top row. }
\end{center}
\end{figure}

In this section, we propose that the $N$-boson realization of the $\W^e_\infty[1]$ algebra, can be extended to provide a realisation of a larger algebra. Indeed, out of $N$ bosons, one can construct not only bilinear terms, but trilinear terms, quartic terms and so forth, invariant under the $S_N$ symmetry,  as is \sct{Review}. The difference, in this case, is that we restrict the operators such that there is no operator with order $p > N$.  

This comes about naturally, if we start with $\W_{1+N}$ generators, as in the top row in Figure~(\ref{rectfig}).  As we saw in the last section, we can choose the $\W_{1+N}$ generators to be either of the form of the elementary symmetric polynomials or of the form of the power sums (or indeed any other symmetric polynomial basis) at $c=N$. The full algebra is generated by derivatives acting on the constituent terms of each $\W_{1+N}$ generator. The full set of operators is the realization of a algebra that we call the $\R\W[N]$ algebra. This construction provides a realization of the $\R\W[N]$ algebra at finite central charge $N$. We show closure of this algebra in \sct{coa}.

To count the total number of generators of this algebra, we point out there is a correspondence that is easily seen (in the power sum basis of \equ{NGP}, for instance,) between the $\W_{1+N}$ generators and the chiral operators of a single boson. Indeed, if we restrict the set of chiral operators for a single boson field such that an operator consists of no more than $N$ $d\phi$, with the number of derivatives acting on the fields unrestricted, there is a one-to-one correspondence between this set and the operators in Figure~(\ref{rectfig}). The total count for this set of operators is the number of ways one can partition an integer (varying from $1$ to $\infty$) into $r$ parts, where $r$ varies from $1$ to $N$. It is therefore given by
\be 
\label{chRWN}
	\prod_{r=1}^{N}\frac{1}{(1-q^r)}\,.
\ee
This then is the count of generators of our proposed $\R\W[N]$ algebra. At finite central charge, one would expect a representation of this algebra to truncate down to a representation of a smaller, as yet unknown algebra. However, we will show in \sct{acc} that at infinite central charge, \equ{chRWN} represents the generating function of the $\R\W[N]$ algebra.

\subsubsection{Closure of algebra \label{coa}}

As in \sct{ff2}, using $N$ bosonic fields one can also construct a realization of the $\W^{e}_{\infty}$ algebra. 
In this $\W^e_{\infty}[1]$ basis, the $\R\W[N]$ algebra is generated by the set of operators that form the $([0^{p-1},1,0,\ldots];0)$ representations of $\W^e_{\infty}[1]$ where $p$ ranges from $1$ to $N$. This is illustrated by the columns in Figure~(\ref{rectfig}). The second column corresponds to the representation $([0,1,0,\ldots];0)$ and the generators of the $\W^e_\infty[1]$ algebra. The commutator of any operator in a given column with a $ \W^e_{\infty}$ generator gives rise to an operator in the same column. In the $\W_{1+N}$ basis, the $\R\W[N]$ algebra is generated by operators that form representations of $\W_{1+N}$.  Furthermore, every $\W_{1+N}$ operator can be expressed as a sum of $\W^e_\infty[1]$ operators (that lie in the rectangle) and vice-versa. 

Using this structure, we can postulate the existence of a closed algebra $\R\W[N]$. Let us outline how to compute the commutators (or corresponding operator product expansions) of any two generators belonging to $\R\W[N]$, which we denote by $\R\W_1$ and $\R\W_2$. The first generator $\R\W_1$ can always be expressed as a descendant of some highest weight state of $\W^{e}_{\infty}$; let us denote this $\W^{e}_{\infty}$ hw state as $\W^{e}_{\textrm{hw}}$.  
Since  $\W^{e}_{\textrm{hw}}$ is also a $\W_{1+N}$ generator, the second operator $\R\W_2$ has a known commutation relation with $\W^{e}_{\textrm{hw}}$, determined by the $\W_{1+N}$ algebra. The operators that one gets on the r.h.s of such a commutation relation will be $\W_{1+N}$ operators that lie in the rectangle.
Once we know the commutator $\[\W^{e}_{\textrm{hw}},\R\W_2\]$, we can reexpress it in the $\W^{e}_{\infty}$ basis and thus find the commutator $\[\R\W_1,\R\W_2\]$, since $\R\W_1$ is a descendant of $\W^{e}_{\textrm{hw}}$. Note that we could have switched $\W_{1+N}$ with $\W^{e}_{\infty}$ in the preceding discussion.

\subsubsection{Deformation to arbitrary central charge \label{acc}}
We now show that this algebra can be deformed to arbitrary integral central charge and that in particular that we can take the central charge to infinity. The currents in \equ{bccurrents} are a realization of the $\W_{1+N}$ algebra at central charge $c$ given in terms of $N$ and the background charge parameter $\alpha_0$ in \equ{bcc}. Clearly, changing the background charge changes the central charge of the theory while keeping the total number of bosons fixed to $N$. For our purpose, however, we are interested in changing the total number of bosons as well, since we want a realization of the $\W^{e}_{\infty}[1]$ algebra at infinite $c$ as well. In other words, we are interested in a $D$-boson representation of the algebra where $D\geq N$. 

This can be achieved as follows \cite{Romans,Lu}. Let us write down the Miura construction for the $\W_N$ algebra:
\begin{equation}
\label{miurar1}
\; : \prod_{j=1}^{N} \Big( \alpha_0 \partial + i \, \e_j \cdot \partial \phi_j(z) \Big) : = \sum_{k=0}^N \W^{(k)}(z) (\alpha_0 \partial)^{N-k}\,.
\end{equation}
Here, we take
\be
	\e_i \cdot \e_j = \delta_{ij} - \frac{1}{N}\, \textrm{~~and~~} \sum_{i=1}^{N} \e_i =0.
\ee
By choosing a particular basis for the $\e_i$ one can write down the $\W_N$ currents explicitly. Here we make the choice in Ref.~\cite{Lu}. 
The stress-energy tensor is then given by
\be
\label{ST2}
	\W^{(2)}= \sum_{j=2}^N \( \frac{1}{2} (\pa \p_j)^2 + \frac{1}{2} \alpha_0  \sqrt{j(j-1)}  \, i \, \pa^2 \p_j \) \,,
\ee
corresponding to the central charge
\be
	c = N -1 -  \alpha_0^2 \, N(N^2-1)\,.
\ee
Since the scalar field $\phi_2$ only occurs via its energy-momentum tensor in \equ{ST2} and in all other $\W_N$ currents (see \cite{Lu} for details), one can replace it with a different energy-momentum tensor with the same central charge
\be
\label{ceff}
	c^{\textrm{eff} } = 1 - 6 \alpha_0^2\,.
\ee 
The total central charge remains
\be
	c = c^{\textrm{eff} } + c^{\textrm{rem} }\,,
\ee
where
\be
	c^{ \textrm{rem} }= N-2 -  \alpha_0^2 \, (N^3 - N - 6)\,.
\ee
The stress-energy tensor is now
\begin{align}
\label{ST3}
	\W^{(2)} &= T^{\textrm{eff} } + T^{\textrm{rem} } \nn \\ &= \sum_{j=1}^{D-N+1} \( \frac{1}{2} (\pa \psi_j)^2 + \frac{1}{2} \alpha_j  \, i \, \pa^2 \psi_j \) +\sum_{j=3}^N \( \frac{1}{2} (\pa \p_j)^2 + \frac{1}{2} \alpha_0  \sqrt{j(j-1)}  \, i \, \pa^2 \p_j \) \,,
\end{align}
with the $\alpha_j $ suitably chosen such that the central charge of $T^{\textrm{eff} }$ is given by \equ{ceff}. This construction, thus, furnishes a representation of the $\W_N$ algebra in terms of a total of $D -1$ bosons.  

The goal of this section is to obtain a realization of the $\W_N$ algebra at $c=\infty$. This can be achieved by taking $\alpha_0^2$ to $-\infty$. As a result we can take $D\rightarrow \infty$, while $N$ remains finite.  Both $c^{\textrm{eff} }$ and  $c^{\textrm{rem} }$ diverge in this case. Because $D\rightarrow \infty$, we can take the $\alpha_j$ to be identically zero. 
As in the construction of \sct{Review}, we start with an infinite number of bosons, but now we are dividing them into two parts: a collection of $\sim N$ bosons and a collection of $D - N$ bosons distinguished by the differing values of the background charge. This enables us to construct a realization of the $\W_N$ algebra using an infinite number of bosons. 

The currents of the $\W_{1+ N}$ algebra can be written in a $\W_N$ basis by adding a spin one current that commutes with all currents of the $\W_N$ algebra. Additionally, one can check that the $\W_{1+N}$ currents given in \equ{bccurrents} can be written in terms of $\W_{1+(N-1)}$ currents and an additional boson field. Applying this recursively, one will end with a realization of the $\W_{1+N}$ algebra in terms of $N-2$ bosons, a spin one current and an arbitrary stress-energy tensor. Hence, the above construction for $\W_N$ works for $\W_{1+N}$ as well. 

We can now extend the algebra by scattering additional derivatives on the constituent terms of the $\W_{1+N}$ currents.  The only change that has occurred because of the change in central charge is an increase in the total number of bosonic fields and additional lower-order terms multiplied by powers of the deformation parameter $\alpha_0$. However, for the total count of operators of the extended algebra, it is the term with the highest order in each $\W_{1+N}$ current which matters, not the total number of bosons. The lower-order terms are irrelevant as well, since they serve as correction terms. Thus the total count of operators is still given by \equ{chRWN}.  

The construction we have outlined above no longer corresponds to the symmetric product orbifold in the large $D$ limit. As we are taking $D\rightarrow \infty$ we do have an infinite number of bosons in total, however in choosing a finite subset of bosons out of this set we have broken the original $S_D$ symmetry. It would be interesting to find a geometrical interpretation of this construction. 

\section{Identities \label{Identities}}
The assertions in \sct{main} can be quantitatively checked by computing the characters of the relevant $\W$-algebras. Indeed, if our assertion is true, the character in \equ{chRWN} should be expressible as a sum of characters of the representations of the $\W$-algebras. We show in this section that this is precisely what occurs because of the existence of non-trivial $q$-series identities. 

\subsection{Vertical identity}
We first show that the $\R\W[N]$  algebra decomposes into $N$ representations of $\W^e_\infty$. For this we use a finite version of the $q$-binomial identity
\be\label{veid} 
(1-q)\prod_{b=1}^{N}\frac{1}{(1-q^b)}  =
 (1-q)\Bigl(1+ \sum_{r=1}^{N}b^{({\rm wedge}) [\lambda=1]}_{([0^{r-1},1,0,\ldots,0];0)} (q) \Bigr) \ .
\ee
where
\be
\label{qveid}
b^{({\rm wedge})[\lambda=1]}_{([0^{r-1},1,0,\ldots,0];0)} (q) = \frac{q^{r}}{\prod_{k=1}^r(1-q^k)} \, .
\ee
This identity is a straightforward truncation of the identity in \equ{VIfull}. 
\subsection{Horizontal identity \label{glid}}
There is a second decomposition of the higher spin rectangle algebra into an infinite number of representations of the wedge algebra of a $\W$-algebra with a finite number of generators. Note that at $c=N-1$, since the level $k \rightarrow \infty$, there are an infinite number of representations of the $\W$-algebra corresponding to the coset: 
\be
	\frac{SU(N)_k \otimes SU(N)_1}{SU(N)_{k+1}}\,.
\ee
These representations can be analytically continued to $c=\infty$ by varying $\alpha_0$ \cite{Gaberdiel:2012ku}. In terms of $N$ and $k$, $\alpha_0$ is given by
\be
	\alpha_0^2 = \frac{1}{(N+k)(N+k+1)}\,.
\ee
To take the $\W_N$ theory to  $c\rightarrow\infty$, we change $\alpha_0^2$ from $0$ to $-\infty$, while the level $k$ varies from $\infty$ to $-N-1$. The theory is non-unitary in this limit.

In terms of characters, the decomposition of the higher spin rectangle algebra is given by
\be\label{hoid} 
\prod_{b=1}^{N}\frac{1}{(1-q^b)}  =
 \Bigl(1+ \sum_{b=1}^{\infty} \P^{\mathfrak{gl}(N+1)}_{[b,0,\ldots,0,b]} (z_i) \Bigr) \ .
\ee
Here, $\P_\L^{\mathfrak{gl}(N+1)}$ is a $\mathfrak{gl}(N+1)$ character. We define it in terms of a $\mathfrak{sl}(N+1)$ character as
\be \label{glnc}
	\P_\L^{\mathfrak{gl}(N+1)} (z_i)= q^{C_2(\L)} \c_\L^{\mathfrak{sl}(N+1)}(\tilde{z}_i)
\ee
where $C_2(\L)$ is the quadratic Casimir for the representation $\L$. The variables $z_i$ are defined as $z_i = q^{i-1}$, where $i$ varies from $1$ to $N+1$ and the variable $\tilde{z}_i$ as $q^{i-(N+2)/2}$. Note that $\prod_{i=1}^{N+1} \tilde{z}_i= 1$. The $\mathfrak{sl}(N+1)$ character is given in terms of Schur polynomials as \cite{Gaberdiel:2010}
\be
	\P_\L^{\mathfrak{sl}(N+1)}(\tilde{z}_i) = q^{-\frac{N+1}{2} B(\L) + \frac{1}{2}\sum_j c_j^2}  \prod_{i=2}^{N+1} \prod_{j=1}^{i-1}\frac{1-q^{(l_j -l_i +i -j)}}{1-q^{(i -j)}}\,.
\ee
Here $B(\L)$ is the total number of boxes in the Young diagram of the representation $\L$. The $l_i$ are defined in terms of the row lengths $r_i\,$, where $i$ varies from $1$ to $N+1$ and $r_{N+1}=0\,$,  of the Young diagram as $l_i=r_i - \frac{B(\L)}{N+1}$.
With these definitions, \equ{hoid} is identical to the following $q$-identity:
\begin{align}
	\prod_{b=1}^N \frac{1}{1-q^b} &=1+ \sum_{k=1}^{\infty} q^{k^2} \frac{(1-q^{N+2k})(1-q^{N-1+k})^2\cdots(1-q^{1+k})^2}{(1-q^{N})(1-q^{N-1})^2\cdots(1-q)^2}\nn \\
	&=1+ \sum_{k=1}^{\infty} q^{k^2} \frac{(1-q^{N+2k})(1-q^{N+k-1})^2\cdots(1-q^{N+1})^2(1-q^{N})}{(1-q^{k})^2\cdots(1-q)^2}\,.
\label{qhoid}
\end{align}
The above identity is a special case of the Rogers-Fine identity (See, {\it e.g.}, Ref.~\cite{Alladi}). The Rogers-Fine identity is in general a three-parameter identity which can be reduced to the following two-parameter identity:
\be
	\frac{(acq)_\infty}{(cq)_\infty}= 1+ \sum_{k=1}^{\infty} \frac{c^k q^{k^2} (a)_k (acq)_{k-1} (1-a c q^{2k})}{(q)_k(cq)_k}\,,
\ee
where $(a)_n \equiv (a;q)_n = \prod_{j=0}^{n-1} (1- a q^j)\,.$  Making the substitutions $c=1$ and $a=q^N$, we recover the identity in \equ{qhoid}.

\subsubsection{Interpretation of the identity}
The wedge group of a $\W$-algebra typically emerges only in the $c\rightarrow  \infty$ limit. Indeed, for the case of the $\W_N$ algebra for the representation $(\Lambda_+;0)$ where $\Lambda_+ = [k,0,\ldots,0,k]$, the character is given by \cite{Perlmutter:2012}
\be
	q^{\frac{N-1-c}{24}}\,q^{O\(\frac{1}{c}\)}\,\frac{ \prod_{s=2}^{N} \prod_{k=1}^{s-1} (1-q^k)}{\eta^{(N-1)}} \, \c^{\mathfrak{sl}(N)}_{\Lambda_+}(\tilde{z}_i)\,,
\ee
in the large $c$ limit. This character reduces to 
\be
	q^{\frac{N-1-c}{24}} \,\c^{\mathfrak{sl}(N)}_{\Lambda_+}(\tilde{z}_i)
\ee
if we restrict to the wedge modes.

The $k=1$ term in the RHS of \equ{qhoid} is the character of the $[1,0^{N-2},1]$ representation of $GL(N+1,C)$. To illustrate for $N=2$, the representation $\Lambda=[1,1]$ has character
\begin{align}
	\P_{[1,1]}^{\mathfrak{gl}(3)} (z_i)&= q + 2 q^2 + 2 q^3 + 2 q^4 +  q^5 \nn \\ 
	&= q^3 \{(\tfrac{1}{q} +1+q)+(\tfrac{1}{q^2} + \tfrac{1}{q}  +1 + q +q^2)\} \,.
\end{align}
Here, the expression in the outer bracket is the character for $SL(3,C)$ corresponding to the wedge modes:
\be 
	L_{-1},L_0,L_{+1}~~\textrm{and}~~\W^{(3)}_{-2},\W^{(3)}_{-1},\W^{(3)}_{0},\W^{(3)}_{+1},\W^{(3)}_{+2}\,,
\ee
of the $\W_3$ algebra. One can construct the $\W_3$ algebra using a two-boson realization, where the spin $3$ current would be order $3$ in the $\partial \phi$. In general, the standard Miura construction uses $N$ bosons to construct the algebra $\W_{(N+1)}$, with the spin of fields ranging from $2$ to $N$ and wedge group $SL(N+1,C)$. To make contact with the bosonic construction in \sct{main}, where the highest order term in the $\W$-algebra currents is of order $N$, we are interested in the algebra $\W_{1+N}$, with spins ranging from $1$ to $N$ and the wedge group $GL(N,C)$. To get this algebra we need to further decompose the group $GL(N+1,C)$ as $GL(N,C)\otimes GL(1,C)$. 

Let us work out the branching rule for the decomposition of the  $[1,0,\ldots,0,1]$  rep of $GL(N,C)$ into these subgroups \cite{Stanley, goodwall}.
The representation $[1,0,\ldots,0,1]$ corresponds to the Young tableau $(2,1,\ldots,1)$, where each entry in the list is the number of boxes in the rows of the Young diagram. Henceforth, in this section, we will use the Young tableau notation to denote a representation. Specifically for the $(2,1)$ rep of $GL(3,C)$, the branching rule is 
\be
	(2,1) = (1)\otimes(2)  \pmb{\oplus}  (1)\otimes(1,1) \pmb{\oplus}  (2)\otimes(1)  \pmb{\oplus}  (0)\otimes(2,1) \,.
\ee
where the first factor in each term on the RHS is the representation of $GL(1,C)$ and the second factor is the representation of $GL(2,C)$. The corresponding character decomposition is
\be
	q + 2 q^2 + 2 q^3 + 2 q^4 +  q^5 = q^2(1+q+q^2)+q^2(q) + q^4(1+q)+(q+q^2)\,. 
\ee
This decomposition structure is preserved at arbitrary $N$. The $(2,1,\ldots,1)$ rep of $GL(N+1,C)$ decomposes as
\be
\label{chardec}
	(2,\underbrace{1,\ldots,1}_{N-1}) = (1)\otimes(2,\underbrace{1,\ldots,1}_{N-2})  \pmb{\oplus} (1)\otimes(\underbrace{1,\ldots,1}_N)  \pmb{\oplus} (2)\otimes(\underbrace{1,\ldots,1}_{N-1}) \pmb{\oplus}  (0)\otimes(2,\underbrace{1,\ldots,1}_{N-1}) \,.
\ee
In this decomposition, the first two representations of $GL(N,C)$ can be combined to give the adjoint representation. This representation, corresponding to the wedge modes of the $\W_{1+N}$ algebra, is always present in the decomposition. In terms of free bosons, the zero modes of the adjoint representation map to the bosonic realization of the $\W_{1+N}$ algebra in \sct{main}.

For $k>1$, there is an analogous branching rule for the rep $[k,\ldots,k]$ of $GL(N+1,C)$ into the reps of $GL(N,C)$. The identity in \equ{qhoid}, therefore, represents the decomposition of the generating character of the $\R\W[N]$ algebra into characters of the representations of $GL(N,C)$, which we interpret here to be the wedge group of the $\W_{1+N}$ algebra at $c=\infty$.

Note that as $N \rightarrow \infty$ the identities in \equ{veid} and \equ{qhoid} tend to \equ{VIfull} and \equ{HIfull} respectively.

\section{At finite central charge \label{finiteN} }
While we have constructed the $\R\W[N]$ algebra at infinite central charge, it is natural to ask whether the $\R\W[\infty]$ algebra has consistent truncations at a finite value of the central charge. There are some indications for this. In particular, the $\W_{1+\infty}[0]$ algebra has a neat truncation at finite $c=N$. 

This assertion follows from the triality symmetry of the $\W_\infty[\l]$ algebra. Due to triality symmetry the following algebras are isomorphic 
\be
	\W_\infty[0] \cong \W_\infty[c+1] 
\ee
at the value of central charge $c$.
Extending this triality to the $\W_{1+\infty}[\l]$ algebra leads to $\W_{1+\infty}[0] \cong \W_{1+\infty}[c]$. For our case, the central charge is $c=N$ for which 
\be
	\W_{1+\infty}[0]  \cong \W_{1+\infty}[N] \cong \W_{1+N,k\rightarrow\infty}\,.
\ee
For the second equality we have used the fact that $\W_{\infty}[N] $ truncates to $\W_{N}$ at $c=N-1$. This is the symmetry algebra of the coset $\frac{SU(N)_k \otimes SU(N)_1}{SU(N)_{k+1}}$ in the $k \rightarrow \infty$ limit. In fact, much before the discovery of triality symmetry for the $\W_\infty$ algebra, the relation $\W_{1+\infty}[0] \cong \W_{1+N,k\rightarrow\infty}$ at $c=N$ was already known \cite{Frenkel:1994, Awata:1994}. It was shown in Ref.~\cite{Frenkel:1994} that `quasi-finite' unitary representations of $\W_\infty[0] $ generically reduce to representations of $\W_{1+N}$. This is because at $c=N$ such representations develop extra null vectors. Quotienting by the submodule generated from these null vectors leads to a irreducible representation of $\W_{1+\infty}[0] $ which is isomorphic to a representation of $\W_{1+N} \equiv \W(\mathfrak{gl}(N))$ with the same central charge.

The $\R\W[\infty]$ algebra is composed of $\W_{1+\infty}[0] $ representations at $c=\infty$. We can analytically continue these representations 
 to finite $c$. As $c$ hits integer values, these representations will truncate down to $\W_{1+N}$ reps.  However, it is not clear what happens to the vertical algebra $\W^e_\infty[1]$ at $c=N$. It is expected that this algebra too will truncate to a smaller algebra, but at present this truncation is unknown. This truncation is not as straightforward as for the $\W_{1+\infty}[0]$ case. 

The $\R\W[N]$ algebra at $c=\infty$ does not appear to arise as a direct truncation of $\R\W[\infty]$, this really depends on how many independent parameters the $\R\W[\infty]$ algebra has. Indeed, it could be possible that there are two independent parameters in the commutation relations of the $\R\W[\infty]$ algebra, in which case, the $\R\W[N]$ algebra can arise as a direct truncation. 

The operators that make up the $\R\W[N]$ algebra at $c=N$ are clearly a subset of the chiral operators of the free boson theory on the symmetric product orbifold $(\TT^4)^N/S_N$. The operators of the symmetric product orbifold theory can be enumerated using the methods of Ref.~\cite{elliptic}. However, it is not clear what the significance of the $\R\W[N]$ algebra is to the theory of a single boson on the finite $N$ symmetric product orbifold.

\section{Discussion \label{dis}}
In this paper we have shown that there exists an algebra $\R\W[N]$, parametrized by the integer $N$, which is similar in structure to the $\R\W[\infty]$ algebra. We provided a realization of the algebra using free bosonic fields. So far, we have not worked out the commutation relations for this algebra, but this should be straightforward for small $N$.

By restricting to the wedge algebra of $\R\W[N]$, one can postulate the existence of the lie algebra $\hsr[N]$. The $\hsr[N]$ algebra reduces to the $\hs^e[1]$ algebra (tensored to a $U(1)$ field) for $N=2$ which is the gauge group of three-dimensional Vasiliev theory with even spin gauge fields and a scalar field with mass $\mu =1$. For $N=\infty$, it is synonymous with $\hsf$ which is believed to be the symmetry algebra of string theory on the $AdS_3$ background in its bosonic incarnation. The existence of such an algebra, therefore, indicates bulk theories that interpolate between Vasiliev theory and string theory. Indeed at least in dimension $D=4$, there is a bulk construction for such theories \cite{Vasiliev:2012}. In view of this, we expect that the $\hsr[N]$ algebra can be constructed in the bulk as a quotient of the universal enveloping algebra of $\hs^e[\lambda]$. It would be interesting to check whether one can consistently couple a matter sector to such theories as for the $\hsf$ algebra \cite{Raeymaekers:2016}.

For the case $N=2$, the corresponding asymptotic symmetry algebra of $\hsr[N]$ is $\W^e_\infty[1]$. The $\W_{1+\infty}[1]$ algebra, for which $\W^e_\infty[1]$ is a subalgebra, has been shown to be isomorphic to the affine Yangian $\mathcal Y$ of $\mathfrak{gl}(1)$ \cite{Prochazka:2015,Gaberdiel:2017b}. There is also a different manner in which $\W_{N}$ algebras are related to Yangians.  While this paper is concerned with `affine' $\W$-algebras, there also exist finite versions of $\W$-algebras, constructed by restricting the affine $\W$-algebras to zero modes. For the case of the usual $\W_N$ algebras, the corresponding finite algebras are abelian algebras. Such finite $\W$-algebras are related to the Yangian of $\mathfrak{sl}(N)$ \cite{Ragoucy:1998}. It is, therefore, plausible that the full $\R\W[N]$ algebra is connected to the affine Yangian of $\mathfrak{gl}(N)$. In fact, although it is expected that $\R\W[\infty]$ is related to the Yangian of some affine algebra, so far, the exact connection has remained elusive. This connection may be easier to see in the simpler case of the $\R\W[N]$ algebra and deserves to be investigated. 

It is likely that the $\hsf$ algebra has other truncations similar to the $\hsr$ algebra but consisting of a different set of operators. In particular, there exist $q$-series identities which indicate this. A different starting point, other than the one examined in this paper, is to consider $Z_N$ parafermionic theories on the symmetric product orbifold and to write down the set of chiral single-trace operators for these theories in the $N\rightarrow \infty$ limit. One would expect the resulting chiral algebra to to be connected to the $\hsf$ algebra, since the $Z_N$ parafermion theory in the $N\rightarrow \infty$ limit reduces to the complex boson theory \cite{Bakas:1990ry}. 

Identities involving $q$-series have played a major role in understanding the chiral algebras in Ref.~\cite{Gaberdiel:2015a} and this paper. In this connection, we would like to point out a slightly modified question from N.J. Fine's text~\cite{Fine:1988} on hypergeometric series: ``Why does the series
\be
	\sum_{n=1}^{N} \frac{ (a q)_n }{(b q)_n}
\ee
have so much structure and and yield such diverse and interesting results in such a natural way?''. The results of this paper and Ref.~\cite{Gaberdiel:2015a} indicate that one answer to this question is: This series has structure because it is the wedge character of the (supersymmetric generalization) of the $\R\W[N]$ algebra, with the identities involving the above expression capturing the relation of this algebra to the $\W$-algebras.

\section*{Acknowledgements}
I am grateful to Rajesh Gopakumar and Yang-Hui He for advice and encouragement. I thank Sanjaye Ramgoolam, Bogdan Stefanski, Alessandro Torrielli, G.M.T Watts and especially Matthias Gaberdiel for discussions. I thank ICTS, Bangalore for hospitality during the course of this work. 

\end{document}